\documentclass[letter,12pt]{article}
\usepackage{graphicx,amssymb,amsmath,epstopdf,color,hyperref}

\def\beq{\begin{equation}}
\def\eeq{\end{equation}}
\def\bea{\begin{eqnarray}}
\def\eea{\end{eqnarray}}

\def\gev{\, {\rm GeV}}

\newcommand{\gsim}{\lower.7ex\hbox{$\;\stackrel{\textstyle>}{\sim}\;$}}
\newcommand{\lsim}{\lower.7ex\hbox{$\;\stackrel{\textstyle<}{\sim}\;$}}

\begin{document}

\begin{flushright}
August 10, 2018
\end{flushright}

\vspace{0.07in}

\noindent
\begin{center}
%{\bf\Large Naturalness, finetuning functionals, \\ and probabilistic interpretations of theories} \\
%{\bf\large Naturalness, Finetuning, and the Likelihood of a Theory}\\
{\bf\large Naturalness, Extra-Empirical Theory Assessments, \\ and the Implications of Skepticism}\\

\vspace{0.5cm}
{ James D. Wells}%\footnote{email: jwells@umich.edu}

{\it Leinweber Center for Theoretical Physics \\
Physics Department, University of Michigan \\
Ann Arbor, MI 48109-1040 USA}\\
%{\tt\small email: jwells@umich.edu}

%\vspace{0.75cm}

%(\today)
\end{center}

%\bigskip\bigskip
\noindent
{\it Abstract:} 
Naturalness is an extra-empirical quality that aims to assess plausibility of a theory. Finetuning measures are often deputized to quantify the task. However, knowing statistical distributions on parameters appears necessary. Such meta-theories are not known yet. A critical discussion of these issues is presented, including their possible resolutions in fixed points. Both agreement to and skepticism of naturalness's utility remains credible, as is skepticism to any extra-empirical theory assessment (SEETA) that claims to identify ``more correct" theories that are equally empirically adequate. The severe implications of SEETA are set forward in some detail. We conclude with a summary and discussion of the viability of three main viewpoints toward naturalness and finetuning, where the ``moderate naturalness position" is suggested to be most appealing, not suffering from the disquietudes of the extreme pro- and anti-naturalness positions.

%\vspace{0.7cm}
%\begin{center}
%{\it Based on lecture delivered at workshop on} \\ 
%{\it ``Naturalness, Hierarchy, and Finetuning", Workshop, Aachen, Germany, 28 February 2018 }
%\end{center}

\vfill\eject

%%%%%%%%%%%%%%%%%%%%%%%%%%%%%%%%%%%%%%%%%%%%%%%%%%%%%%%%%%%%%
\section{Extra-empirical attributes of theories}

{\it Empirical adequacy}\,\footnote{Italicized words appearing in this article are defined in more detail in~\cite{Wells:Lexicon}.} is the pre-eminent requirement of a {\it theory}. However, a key problem in science is the underdetermination of theory based on observed phenomena. Many theories, in fact an infinite number of theories, are consistent with all known observations. This assertion may be qualitatively true, but assuredly it is technically true when we realize that there are a finite number of observations made to compare with theory, and all observations have uncertainty (e.g., the mass of the $Z$ boson is $M_Z=91.1876\pm 0.0021\gev$~\cite{PDG16}). 

Faced with a large number of {\it concordant} theories (i.e., equally empirically adequate), one looks to additional extra-empirical criteria to further refine assessments of their value. These extra-empirical attributes to {\it theories} include {\it simplicity}, {\it testability}, {\it falsifiability}\footnote{For future discussion, the distinction between ``testable" and ``falsifiable" theories will be important. A ``testable" theory is ``one that contains at least one point in parameter space
that is capable of yielding evidence for new physics beyond the standard theory in future
experiments," whereas a ``falsifiable theory" is one ``whose entire parameter space could conceivably
be ruled out (i.e., shown to be non-concordant) by a specified collection of experiments
and analysis that can be done in the future"~\cite{Wells:Lexicon}. A falsifiable theory is a testable theory, but a testable theory might not be falsifiable.}
, {\it naturalness}, {\it calculability}, and {\it diversity}. None of these attributes has been proven to be logically necessary for an {\it authentic} theory\footnote{In~\cite{Wells:Lexicon} an {\it authentic theory} is defined to be ``... one that has a point in its parameter space ... that is concordant with any conceivable experiment that
could possibly be performed in the theory's domain of applicability." More colloquially, it is the ``correct theory" or, in the case of a less ambitious effective theory, a ``more correct theory" or ``deeper theory."}, although an {\it authentic} theory may possess them. However, there still may be important reasons to evoke these considerations in the pursuit of scientific progress. For example, a scientist may wish to make a discovery in his/her lifetime, in which case {\it promptly testable} theories are more important to work on than theories judged to be more likely but not {\it promptly testable}.  Or, a scientist may wish to widen her vision of observable consequences of {\it concordant theories} in order to cast a wider experimental net, which would lead her to pursue {\it diverse theories} over {\it simple theories}.

Another preference might be to identify the subset of {\it concordant theory(ies)} most likely to be {\it authentic} among the much larger collection of concordant theories. Simplicity, falsifiability, calculability, etc., are all not reliable guides to answer this question. However, among the extra-empirical attributes {\it naturalness}~\cite{Giudice:2008bi,Fichet:2012sn,Farina:2013mla,Tavares:2013dga,Kawamura:2013kua,deGouvea:2014xba,Williams:2015gxa,Wells:2016luz,Giudice:2017pzm,Hossenfelder:2018ikr} is the one that most directly speaks to this goal. If the naturalness of a theory has any value at all, it is because it appeals to our quest to sort concordant theories into likely and unlikely candidates -- natural and unnatural theories.

%%%%%%%%%%%%%%%%%%%%%%%%%%%%%%%%%%%%%%%%%%%
\section{Finetuning functional}

In practice, naturalness is often closely tied to notions of finetuning. A theory is unnatural if its parameters require a high degree of finetuning to match observables or other parameters when matching across effective theories (EFTs), and a theory is natural otherwise. One first constructs a finetuning functional $FT$ on a theory $T$ to map the theory to a number $FT[T]$, which is its finetuning. For example, let us denote a set of parameters of the theory by $x_i$, and a set of observables by ${\cal O}_k$. A candidate finetuning functional could be~\cite{Ellis:1986yg,Barbieri:1987fn}
\beq
\label{eq:ftf}
FT[T]={\rm max}\sum_{ik}\left| \frac{x_i}{{\cal O}_k}\frac{\partial{\cal O}_k}{\partial x_i}\right|.
\eeq
The higher the value of $FT[T]$ the more finetuned it is and the less likely it is to be natural, according to this algorithm. For example, one could decide that a theory is natural, or ``FT-natural", only if $FT[T]<F_N$, for some critical finetuning value of $F_N$. 

As the reader may note there are many choices and assumption made beyond identifying the theory to assess whether a theory is FT-natural. In order to construct a finetuning functional one must begin by choosing a  parametrization of the theory, which is not unique. For example, the many different schemes of renormalizable theories (MS-bar, on-shell, etc.) lead to different parametrizations, and in many non-renormalizable theories completely different basis sets of operators are possible through manipulations of the equations of motion. Starting with a particular basis with particular coefficients  it is often possible to go to another basis where coefficients of operators are zero, which may look to the unsuspecting as a magical finetuning.

In addition, the finetuning functional is applied to observable(s). The list of observables in a theory is infinite. Even if we limit ourselves to so-called counting observables, there are an infinite number. For example, there are an infinite number of kinematic configurations of $e^+e^-\to \mu^+\mu^-$ that are rightly classified as observables. Of course, if there are $n$ parameters of the theory then it is usually possible to pick $n$ observables such that all other observables then derive from those. However, the difficulty is choosing which $n$ observables to use in the finetuning functional. An algorithmic method to choose which $n$ observables given some arbitrary theory would only give false comfort in the face of arbitrariness. 

To illustrate further arbitrariness, let us suppose we are allowed to redefine an observable as a function of observables. In that case, we can always construct observables with unit finetuning. The method to do this is to first begin with a set of $n$ parameters $\{ x_i\}$, and a set of $n$ observables $\{  y_i\}$. Then,
\beq
y_i=f_i(x_1,x_2,\ldots,x_n)~~~i=1,2,\ldots,n
\eeq
If we invert these equations we can obtain the parameters in terms of the observables
\beq
x_i=f^{-1}_i(y_1,y_2,\ldots,y_n)~~~i=1,2,\ldots, n.
\eeq
We now define a new set of observables that are
\beq
\hat y_i=f^{-1}_i(y_1,y_2,\ldots,y_n)~~~i=1,2,\ldots, n.
\eeq
With these observables we find that finetuning is always unity
\bea
FT& =&\left| \frac{x_i}{\hat y_k}\frac{\partial \hat y_k}{\partial x_i}\right|
=\frac{x_i}{ (x_k)} \frac{\partial (x_k)}{\partial x_i} \\
&=& \frac{x_i}{x_k}\,\delta_{ik}=1
\eea

Let us now suppose that such constructions of redefined variables are not allowed and that there is an algorithm to choose observables and parameters that is {\it a priori} agreed upon to assess finetuning. Following such a procedure is analogous to experimental searches that search the data for new physics signatures using pre-defined cuts or procedures. It is usually bad practice to bin the data after it is accrued with the sole purpose of maximizing a statistical anomaly, or minimizing it. One decides before the analysis and opens up the box to see if there are anomalies.

However, even with such an approach there are still concerns. For example, let us suppose we have an observable $y$ that depends on the input parameter $x^n$ according to $y=x^n$, where $n$ is some integer power. In this case, the higher the value of $n$ the more finetuned the theory. 
However, it is rather transparent that different powers of $n$ do not affect naturalness as one would intuit. The trouble is that 
this finetuning value is the same no matter what value of $x$ the theory provides, and therefore no matter what the value of the observable $y$. 
Yet, if the theory provides a value of $x$ then there simply is a value of $y$ that comes out. This problem with the finetuning measure was recognized by 
 Anderson \& Casta\~no \cite{Anderson:1994dz,Anderson:1994tr} who attempted to fix the measure by stating a new finetuning functional needs to be defined which is the old finetuning functional divided by the average finetuning. In the $y=x^n$ case, the average finetuning is $n$, since it is the same over all values of $x$ and thus the new finetuning function returns a finetuning of $1$ ($n$ divided by average finetuning of $n$) for all possible values  of $n$. This appears to be going in the right direction, and in cases like this is responsive to our intuitions about finetuning.
 
However, the Anderson, Casta\~no finetuning functional has its own difficulties. For example, most examples are not as simple as the one above -- they do not have a constant finetuning for any input parameters $x_i$. And when the finetuning changes over values of the parameters $x_i$ then one has no choice but to introduce a probability measure over $x_i$ in order to get an ``average finetuning." Such a probability measure introduces yet another extra-empirical assumption about the theory, turning it into a meta-theory, which further calls into question the finetuning functional used to assess naturalness. In addition, since finetuning itself is not firmly rooted in probability theory, adding a sub-component of the functional that introduces probabilities over parameters introduces the burden of justifying probabilities for part of the calculation and then  inexplicably abandons them when applying the finetuning calculation. 

Let us now give an example of how the finetuning functional, even corrected according to Anderson and Casta\~no, can be at odds with probabilistic intepretations  of observable measurements. In many theories, high finetuning results from the cancellation of two large numbers. Let us represent the observable by $y$ and the two large parameters as $x_1$ and $x_2$. The concern is over finetuning for $y=x_1-x_2$. In the Standard Model such an algebraic equation is used to discuss the cancellation of a bare mass term of the Higgs boson with a large cut-off dependent quadratic divergence $\Lambda^2$. Or, more physically, the renormalized SM Higgs mass parameter with a large one-loop correction from an exotic singlet scalar or vector-like quarks~\cite{deGouvea:2014xba,Wells:2016luz}. In supersymmetric theories, the condition for successful electroweak symmetry breaking has the form $y=x_1-x_2$ where $y=m_Z^2$ and $x_1\propto m_{H_u}^2$ and $x_2\propto m_{H_d}^2$ (see, e.g., eq.\,8.1.12 of~\cite{Martin:1997ns}). $m_{H_u}^2$ and $m_{H_d}^2$ are supersymmetric breaking parameters which in many theories of supersymmetric are correlated with superpartner masses of the quarks, leptons and gauge bosons. Since high-energy collisions of LHC have not found superpartners it is expected that these masses are greater than $\sim 1.5\, {\rm TeV}$. This implies that $m_{H_u}^2$ and $m_{H_d}^2$ may be greater than a TeV and thus much larger than $m_{Z}$. 

In all these cases, the concern becomes requiring $y=x_1-x_2$ when $x_1,x_2\gg y$. It looks like a ``finetuned cancellation". Our intuition even suggests that it would be odd that a very large value of $x_1$ would cancel with a very large value of $x_2$ to give me a small value of $y$. The finetuning functional puts a number on that intuition:
\beq
\label{eq:FTx1x2}
FT=\frac{1}{2}\sum_{i=1}^2 \left|\frac{x_i}{y}\frac{\partial y}{\partial x_i}\right|=\frac{1}{2}\left| \frac{x_1+x_2}{x_1-x_2}\right|=\frac{\bar x}{y}
\eeq
where $\bar x$ is the average of $x_1$ and $x_2$. If we applied the Anderson-Casta\~no procedure to this we would have to determine what the probability distributions are of $x_1$ and $x_2$, then find the average $FT$ value and then divide the above equation by that. Thus, it would only change the above equation by a constant and not qualitative change the conclusion that very small values of $y$ compared to much larger values of $x_1$ and $x_2$ are finetuned and therefore unnatural. In the next section we look at this example from the point of view of probability.

%%%%%%%%%%%%%%%%%%%%%%%%%%%%%%%%%%%
\section{Finetuning and probability}

For the case of $y=x_1-x_2$, how rooted in probability and viability is such a finetuning measure as given by eq.~\ref{eq:FTx1x2}? In order to assess this we must assume some probability measure over $x_1$ and $x_2$. If we assume some reasonable probability over these variables and show that FT assessment is incompatible with probability assessments of likely outcome for $y$, we have further reason to suspect the finetuning functional is not a good quantitative measure to assess naturalness. Let us assume that $x_1$ and $x_2$ are flatly distributed over the range of $0$ to $1$. Now, we will approach the question of how likely is a small value of $y$ ($y\ll 1$)  from two points of view.

\begin{figure}[t] 
\begin{center}
\includegraphics[width=0.6\textwidth]{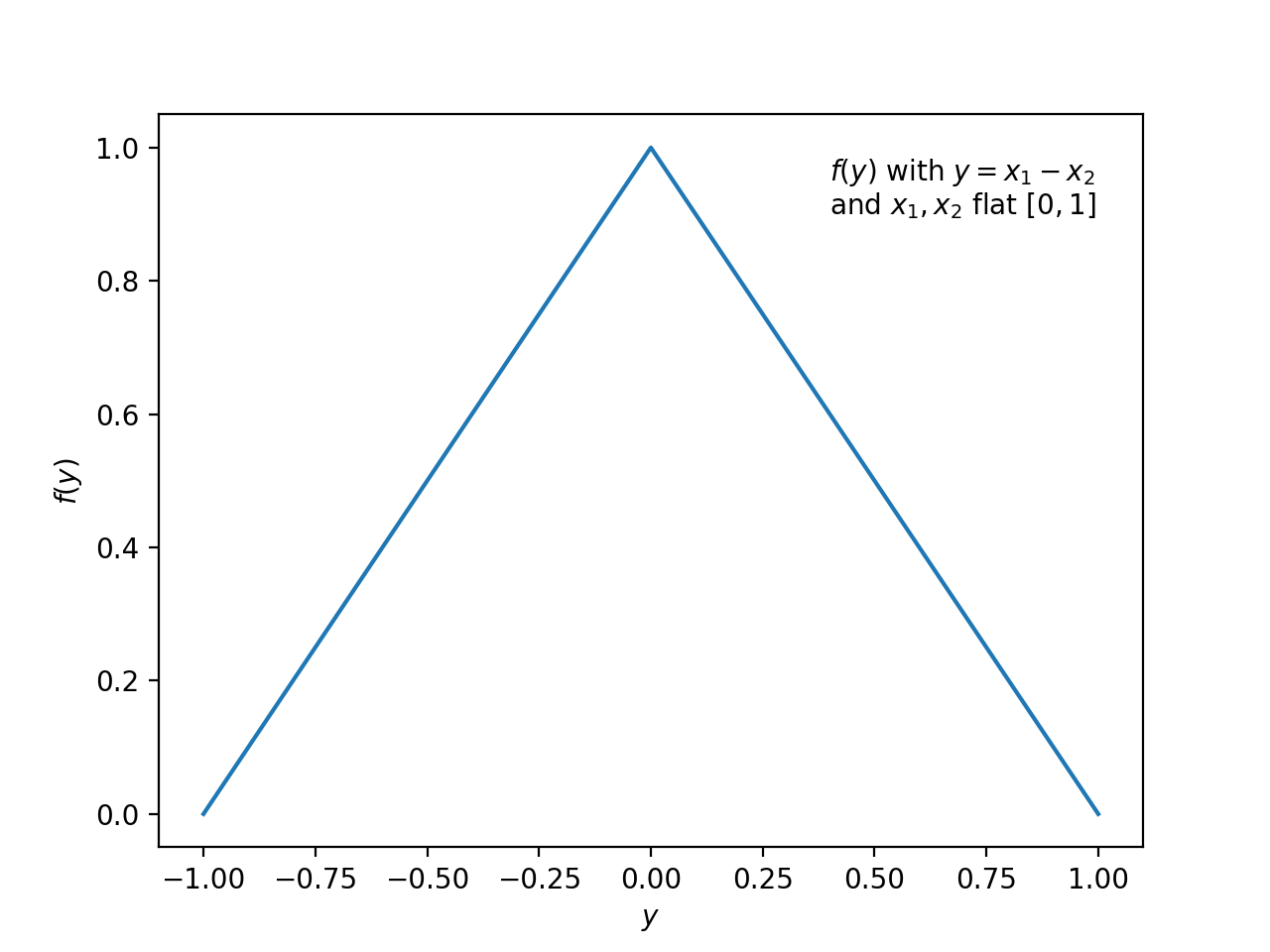} 
\caption{Probability density function $f(y)$ for $y=x_1-x_2$ where $x_1$ and $x_2$ are flatly distributed from 0 to 1. The peak of $f(y)$ is at $y=0$, which according to one interpretation calls into question the claim that small values of $y$ in this case are unnatural.}
\label{fig:fy}
\end{center}
\end{figure}

First, we can ask, given the probability distributions of $x_1$ and $x_2$ is a particular outcome for $y$ improbable. Such questions are often asked in statistics and the first thing one does is finds the probability density function for $f(y)$ given the probability density functions for $f_1(x_1)$ and $f_2(x_2)$. When $x_1$ and $x_2$ are independent random variables the result is a convolution
\beq
f(y)%=\int f_1(y-x)f_2(x)dx
=\left\{ \begin{array}{rl} 1+y & {\rm for}~-1<y<0,\\ 1-y & {\rm for}~0<y<1\\ 0 & {\rm otherwise}\end{array}\right.
\eeq
which is depicted in fig.~\ref{fig:fy}. From the probability distribution $f(y)$ there appears to be no compelling argument that the ``finetuned" values of $y\ll 1$ are improbable. On the contrary, such low values of $y$ are the most probable outcomes. Measurements of $y$ near zero would be near the peak of the probability density of $y$. If we ask what central value of $y_0$ would give the largest probability after integrating over some range of $\Delta y$, we would have to answer $y_0=0$. From this perspective it is hard to say that tiny values of $y$ are improbable, unless somebody employs the specious argument that any exact value of $y$ is improbable, which would inappropriately rule out all values of $y$.

It could be argued that considering equal flat distributions for $x_1$ and $x_2$ from 0 to 1 is not applicable analogy for realistic discussions of  finetuning in physical theories, and it should be a different distribution. Perhaps the distributions should be narrow and far away from zero. Perhaps one should be skewed very differently than the other. There are many possibilities, all of which are {\it a priori} possible in this discussion. The point of the above illustration is to show that one's view of finetuned cancellations  depends on one's assumptions about the distributions, and it is incumbent upon the practitioner to articulate clearly the finetuning measure's connection to probability. Such descriptions and qualifications may even be posited and then tested for how well they conform with data requirements and naturalness intuitions, such was partially done in the discussions on ``focus point supersymmetry"~\cite{Feng:1999mn,Feng:1999zg}.

\begin{figure}[t] 
\begin{center}
\includegraphics[width=0.6\textwidth]{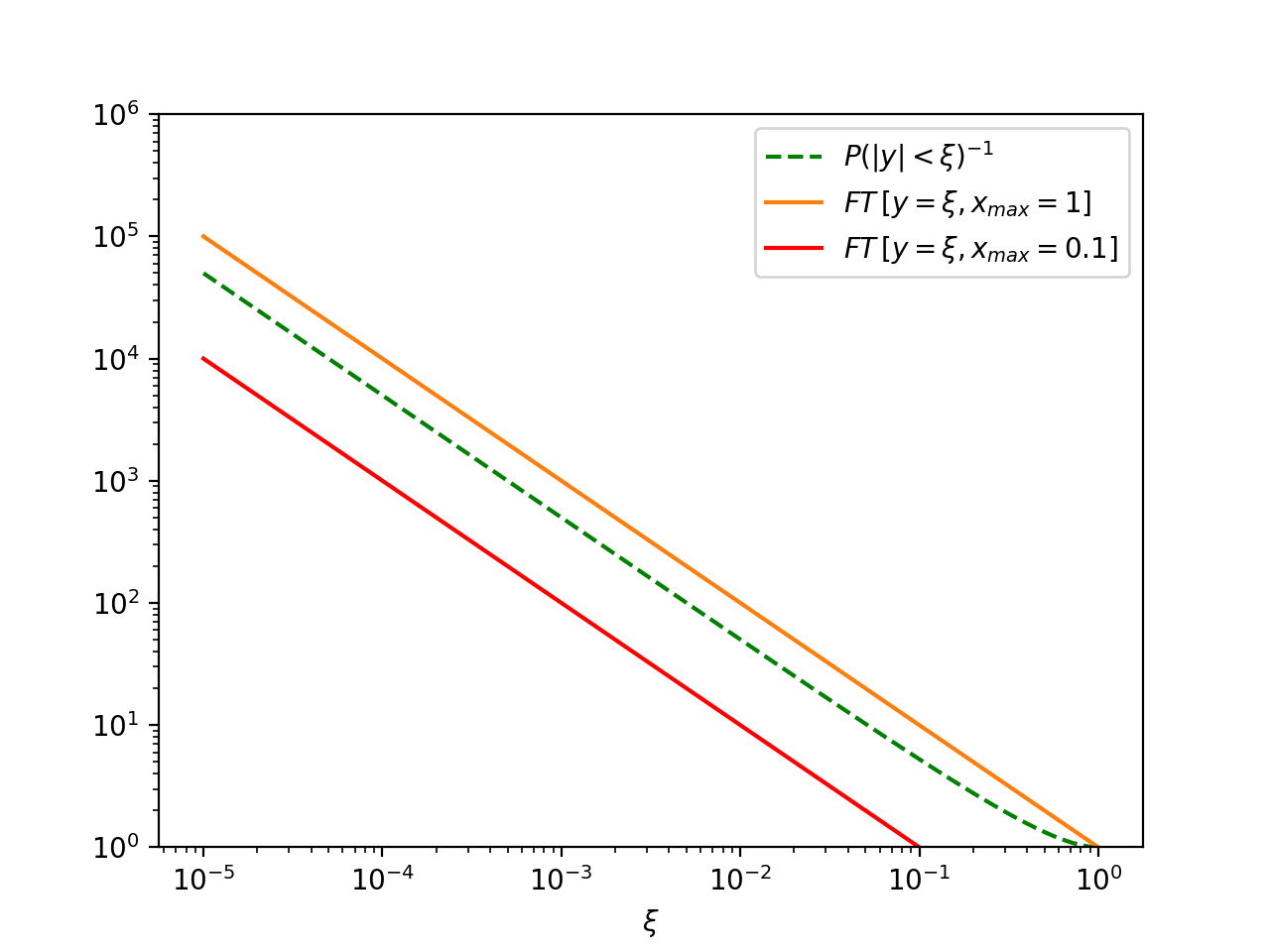} 
\caption{Finetuning computation and inverse probability of $|y|<\xi$ when $y=x_1-x_2$ with $x_1$ and $x_2$ flatly distributed from 0 to 1. The inverse probability of achieving very low values of $|y|$ is correlated well, but not one-to-one, with finetuning in this example. $x_{\rm max}$ is defined to be ${\rm max}(x_1,x_2)$ in the computation for finetuning. The larger the $x_{\rm max}$ the higher the finetuning to achieve low $|y|$.}
\label{fig:ybelow}
\end{center}
\end{figure}

A second way to view this problem is to ask what is the probability of obtaining a value of $|y|$ below some small chosen value $\xi$ given a probability distribution of $x_1$ and $x_2$ near a critical point $y=0$. In fig.~\ref{fig:ybelow} the inverse of that probability is plotted (dashed green line). For example, if we assume $|y|<\xi$, the probability to achieve it is $\sim 2\xi$ for $\xi\ll 1$.  Such a measure yields results similar to traditional finetuning measures. However, plotting the finetuning functional value on this plot requires know $x_{\rm max}={\rm max}(x_1,x_2)$, since finetuning depends on $\bar x$. For example, if $x_{\rm max}=1$ (upper yellow line), then it is even more finetuning to have $|y|<\xi=10^{-3}$ than if $x_{\rm max}=0.1$ (lower red line).  Thus, the FT functional does not return a unique mapping from probability for $|y|<\xi$ but at least it returns that higher finetuning means lower probability, and thus has some probabilistic correlation. 

Despite the more faithful matching of the second algorithm to compute probabilities (cf.\ fig.~\ref{fig:ybelow}) than the first algorithm (cf. fig.~\ref{fig:fy}), it is unclear that the  second algorithm is more appropriate than the first. It may be just as justified (i.e., arbitrary) to ask what the probability is of $|y|\geq \xi$ or $\xi/2<|y|<2\xi$ as it is to ask $|y|\leq \xi$, unless one invokes an additional principle that being close to a ``critical" point in the theory is the reference point where we should compute probabilities. In our case, $y=0$ is the interface between symmetry breaking ($y<0$) and non-symmetry breaking ($y>0$) in a Higgs potential.

Another way of seeing that the second approach could fail at times is to image that the distributions of $x_1$ and $x_2$ yield zero probability density for $y=0.013$ but a significant probability density for $|y|\leq 0.013$. The second approach of attaching viability significance to $y=0.013$ based on computing the probability that $P(|y|<0.013)$ clearly fails, whereas asking for the value of $f(y=0.013)$ to compare with the full probability density function of $y$ is defensible. However, it may only be defensible when probability is zero.

\begin{figure}[t] 
\begin{center}
\includegraphics[width=0.6\textwidth]{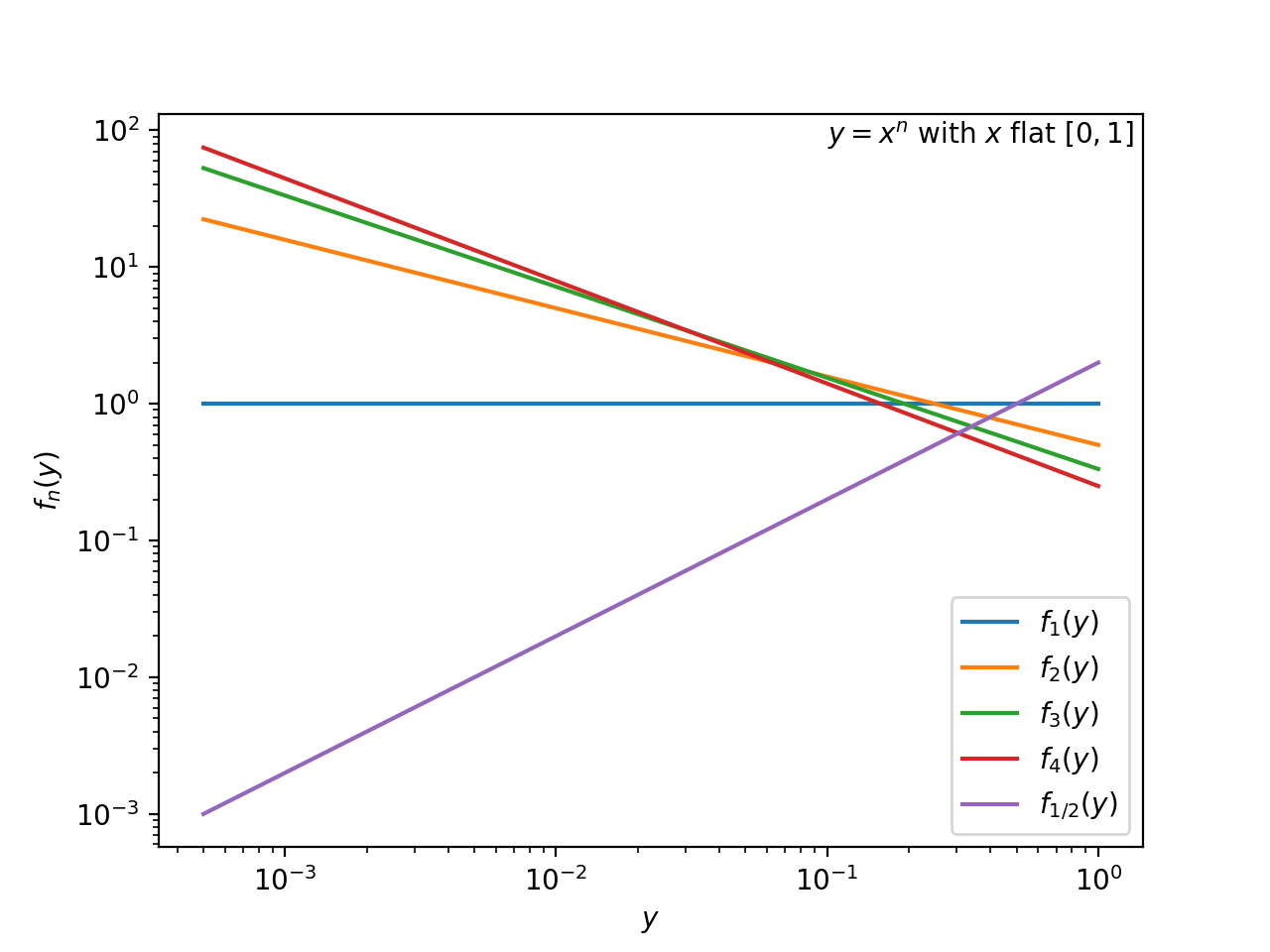} 
\caption{Probability density function $f(y)$ for $y=x^n$ for different values of $n$ and for $x$ flatly distributed from 0 to 1.}
\label{fig:fyxn}
\end{center}
\end{figure}

Let us return to the example earlier of the observable $y$ depending on parameter $x$ through $y=x^n$. We had stated earlier that it was nonsensical that naturalness should depend on $n$, and then found that Anderson-Casta\~no rectified this problem by stating that the finetuning functional should be divided by the average value of finetuning. In that case, $FT=1$ for all choices of $x$ and for all choices of $n$. However, if we put a probability density on $x$ and ask about probability of $y$ rather than finetuning of $y$, we get a very different answer. Fig.~\ref{fig:fyxn} depicts the probability density functions $f_n(y)$  for various values of $n$ given a flat distribution of $x$ from 0 to 1. We see that for high powers of $n$ the probability for lower values of $y$ greatly increases. The finetuning functional does not reveal this issue. 

It is interesting to revisit what we discussed above about redefining sets of observables with respect to this specific case. If we are allowed to construct new observables $\hat y_i$ as a function of some canonical set of independent observables $y_i$, as we discussed above, $\hat y_i=f^{-1}_i(y_1,y_2,\ldots,y_n)$ we can always make a construction where $FT=1$. In this specific case with $y=x^n$, one can redefine $\hat y=y^{1/n}$ and thus $\hat y=x$ and $FT=1$. Furthermore, the probability density of $\hat y$ is flat from 0 to 1. This calls into question the FT measure, but it does not call into question the probability measure. A flat probability for $\hat y$ is equally meaningful to assess the viability of the theory as is the skewed and peaked probability distribution for $y$. However, the ability to redefine observables as functions of observables to get different FT values would mean loss of utility since it is not grounded in assessing probabilities.

%%%%%%%%%%%%%%%%%%%%%%%%%%%
\section{Probability and naturalness}

After considering finetuning and the criticisms of the finetuning functional as a quantitative means to assess naturalness, we have been led to notions of probability distributions on parameters. If naturalness has any connection to the extra-empirical judgments of theory plausibility, which is surely what naturalness discussion is after, we have no recourse but to introduce probability measures across parameters of the theory. A corollary to this is that any discussion about naturalness that does not unambiguously state assumptions about the probability distributions of parameters of the theory cannot be rooted in probability theory and therefore has little to do with theory plausibility.

Now, any explicit claims about probability distributions of parameters is highly controversial, and appears to go beyond the normal scientific endeavor. However, we can discuss normal science within the language of probability distributions on parameters. For example, normal tests of a theory can be reinterpreted as an assumption of $\delta(x_i-x_i^{\rm th})$-function distributions on parameters, which then can be used to compute the distribution of the observables through $y_i^{\rm th}(x_1^{\rm th},x_2^{\rm th},\ldots)$. In the limit of perfect errorless theory calculation the probability distribution of the observables would also be $\delta$-functions: $f(y_i^{\rm th})=\delta(y_i-y_i^{\rm th})$. 

If there is theory uncertainty in the calculation, e.g.\ from finite order perturbation theory, the $\delta$-function distribution of parameters become a somewhat spread-out probability distribution for the observables, whose distribution (and therefore uncertainty) is hard to know, but the standard deviation might be estimated to be the difference in values obtained for the observables when varying the renormalization scale by a factor of two below and above the characteristic energy of the observable process (e.g., $m_b/2<\mu <2m_b$ in $b$-meson observables calculations). But that just complicates the discussion unnecessarily, so let us go back and assume that theory is perfect and we start with a $\delta$-function distribution on parameters and obtain a $\delta$-function distribution of observables, and we compare those observables to the data, which often Gaussian distributed. 

Next, fancy statistical tests are done to see if the predictions are compatible with the measurements, and if so the theory is said to agree with the data and the choice of parameters is then declared acceptable. One does this many times over small changes in the arguments of the $\delta$-function distributions on parameters and finds the full space of parameters where theory predictions are in agreement with observables. In the limiting case that the theory is a valid one, and theory is calculated perfectly, and experiments yield exact results, the $\delta(x_i-x^{\rm th})$-function distributions of parameters yield a $\delta(y_i-y_i^{\rm th})$ distribution of observables that exactly match the data with $y^{\rm th}_i=y^{\rm expt}_i$.

The concept of naturalness invites the theoretician to go beyond this procedure. In the language above, it invites us to consider a distribution of input parameters that are not $\delta$ functions but something more complicated with finite extent. Perhaps the parameters are flatly distributed, or Gaussian distributed, skew-distributed, or something even more complicated. Either way, a choice on distributions must be made or recognized somehow, and then a probability assessment on some outcome must be made. The questions then proliferate: What parameters do I attach probability distributions to? How does one determine which probability distributions are appropriate? What outcomes (observables, parameters?) do I check for probable or improbable? How is probable vs.\ improbable demarcated? All of these questions must be addressed in one form or another. Failure to do so renders any naturalness discussion fuzzy with diminished meaning. At the same time, answering these questions is a highly speculative endeavor given our current understanding of quantum field theory and nature. This is what ultimately may bar naturalness from meaningful technical discourse, even if its qualitative usefulness could be established. Nevertheless,  it is useful to press forward to see if there are qualitative lessons one can learn about theories and their relative degree of probability or improbability compared to other theories.

The above discussion further begs the question of why there should be a probability distribution on parameters at all. Well, there is the formulation of normal science discussed above which is already in terms of probability distributions on parameters, which are $\delta$-functions. And so, a commitment to a particular distribution function has already been made implicitly from this perspective. The question becomes then whether there is another distribution function that is more appropriate. In Landscape discussions it is plausible that parameters are the result of a random choice among a semi-infinite number of solutions from a more fundamental theory. If our universe is one random choice out of these infinite ones we certainly require that the probability of that choice be non-zero. That is not controversial. What becomes more controversial is that we might even wish to demand that the probability of our universe's choice of outcomes be ``generic." In other words, we may wish to require that the value of the joint probability density function over outcomes is not atypically tiny.

%%%%%%%%%%%%%%%%%%%%%%%%%%%%%%%%
\section{Probability flows of gauge couplings}

We have already looked at both the finetuning interpretations and probability interpretations of the cases $y=x^n$ and $y=x_1-x_2$ for $x_i$ inputs and $y$ output. One of the conclusions from that discussion is that finetuning assessments might be useful for judging the plausibility of a theory but only if they match a coherent probability interpretation, and probability interpretations are only possible when distributions on parameters are specified. Let us now proceed to investigate the IR implications of a probability distribution on a UV value of gauge couplings.

First, let us explore the probability flow of QCD gauge coupling. We begin by assuming a flat probability distribution of MS-bar $g_3(M_H)$ coupling at $M_H=10^{15}\gev$ with values from $0$ to $0.6$, chosen such that $g_3(Q)\leq \sqrt{4\pi}$ (``finite") down to $M_Z$. To be clear, this is an ansatz whose implications will be explored. For simplicity we only consider the one-loop $\beta$ function for the renormalization group evolution of $g_3$, which is
\begin{equation}
\frac{dg_3}{d\log Q}=-\frac{b}{2}g_3^3,~~~{\rm where}~b=\frac{7}{8\pi^2}.
\end{equation}
The solution to this equation is
\begin{equation}
\label{eq:g3Q}
g_3(Q)=\frac{g_3(M_H)}{\sqrt{1-g_3(M_H)^2b(Q)}},~~~{\rm where}~b(Q)=b\ln(M_H/Q).
\end{equation}
Note, $b(Q)$ monotonically increases as $Q$ flows to the IR.

Eq.~\ref{eq:g3Q} is analogous to our $y=x^n$ equation given earlier, where $y=g_3(Q)$ and $x=g_3(M_H)$. We know (or rather posit) the distribution on $x$ and we want to know the resulting distribution on $y$. Let us compute the distribution function for $g_3(M_Z)$ assuming the above-stated flat distribution on $g_3(M_H)$. The probability density function is 
\bea
\label{eq:fg3mz}
f(g_3(M_Z))=\frac{1}{\sqrt{4\pi}}\, \frac{(1+4\pi\, b(M_Z))^{1/2}}{(1+g_3(M_Z)^2\,b(M_Z))^{3/2}}.
\eea
If we had chosen $g_3(M_H)$ to be flat from 0 to $\sqrt{4\pi}$ rather than flat from $0$ to $0.6$ then 83\% ($=1-0.6/\sqrt{4\pi}$) of the probability distribution would have flowed to a divergent value of $g_3(M_Z)$. In that case the probability density function could be represented by 
\beq
f'(g_3(M_Z))=0.17\, f(g_3(M_Z))+0.83\,\delta(g_3(M_Z)-\sqrt{4\pi})
\eeq
where we have let $g_3(M_Z)=\sqrt{4\pi}$ be the value of $g_3(M_Z)$ where all the probability now resides for divergent coupling. This is a complication that can be handled, but it is avoided by the original assumption that $g(M_H)$ is flatly distributed from $0$ to $0.6$, and our probability density function $f(g_3(M_Z))$ of eq.~\ref{eq:fg3mz} holds.

\begin{figure}[t] 
\begin{center}
\includegraphics[width=0.6\textwidth]{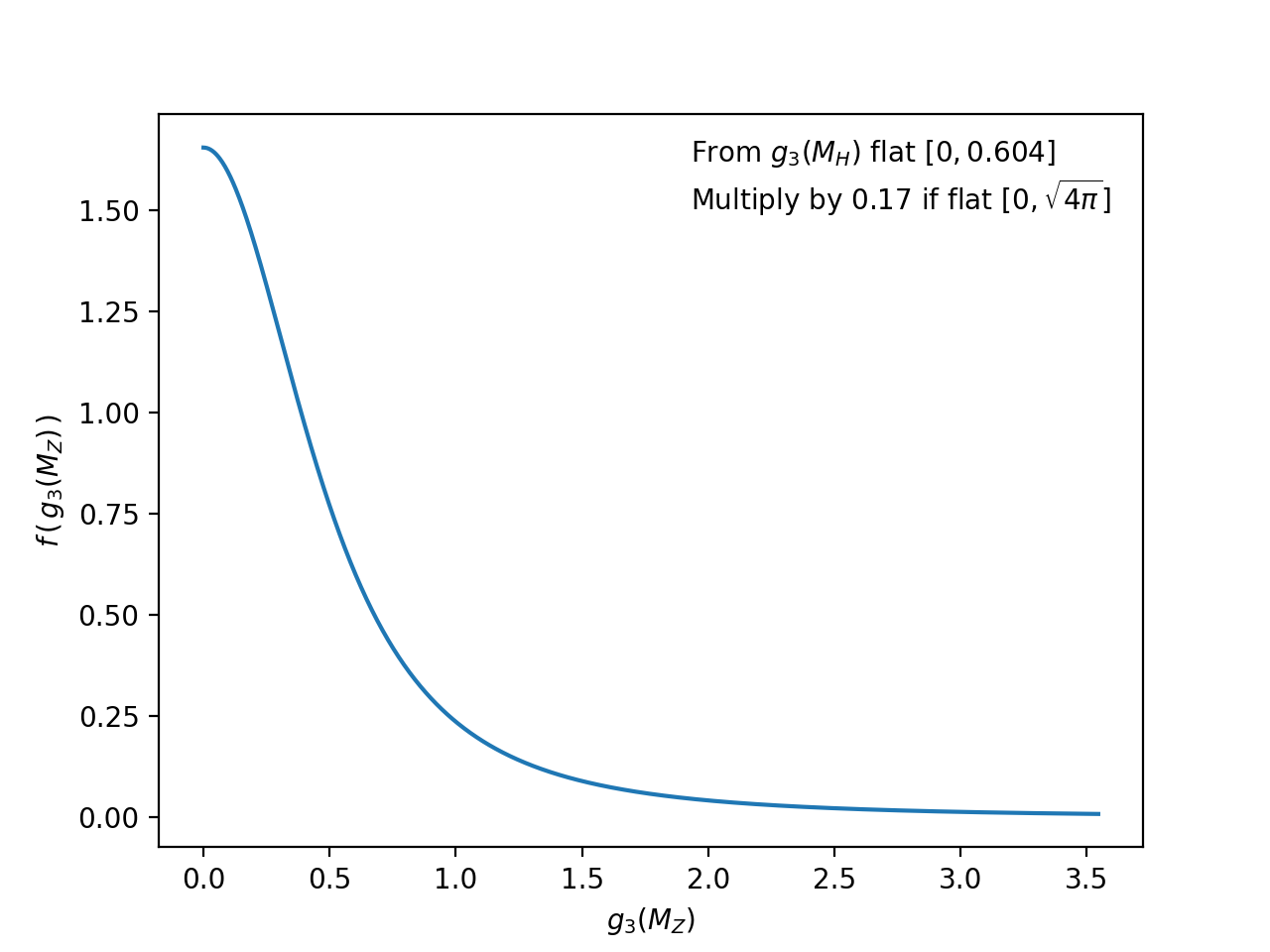} 
\caption{The probability density function of $g_3(M_Z)$ if $g_3(M_H)$ at $M_H=10^{15}\, {\rm GeV}$ is assumed to be flatly distributed from 0 to $\sqrt{4\pi}$.}
\label{fig:fgmz}
\end{center}
\end{figure}

\begin{figure}[t] 
\begin{center}
\includegraphics[width=0.6\textwidth]{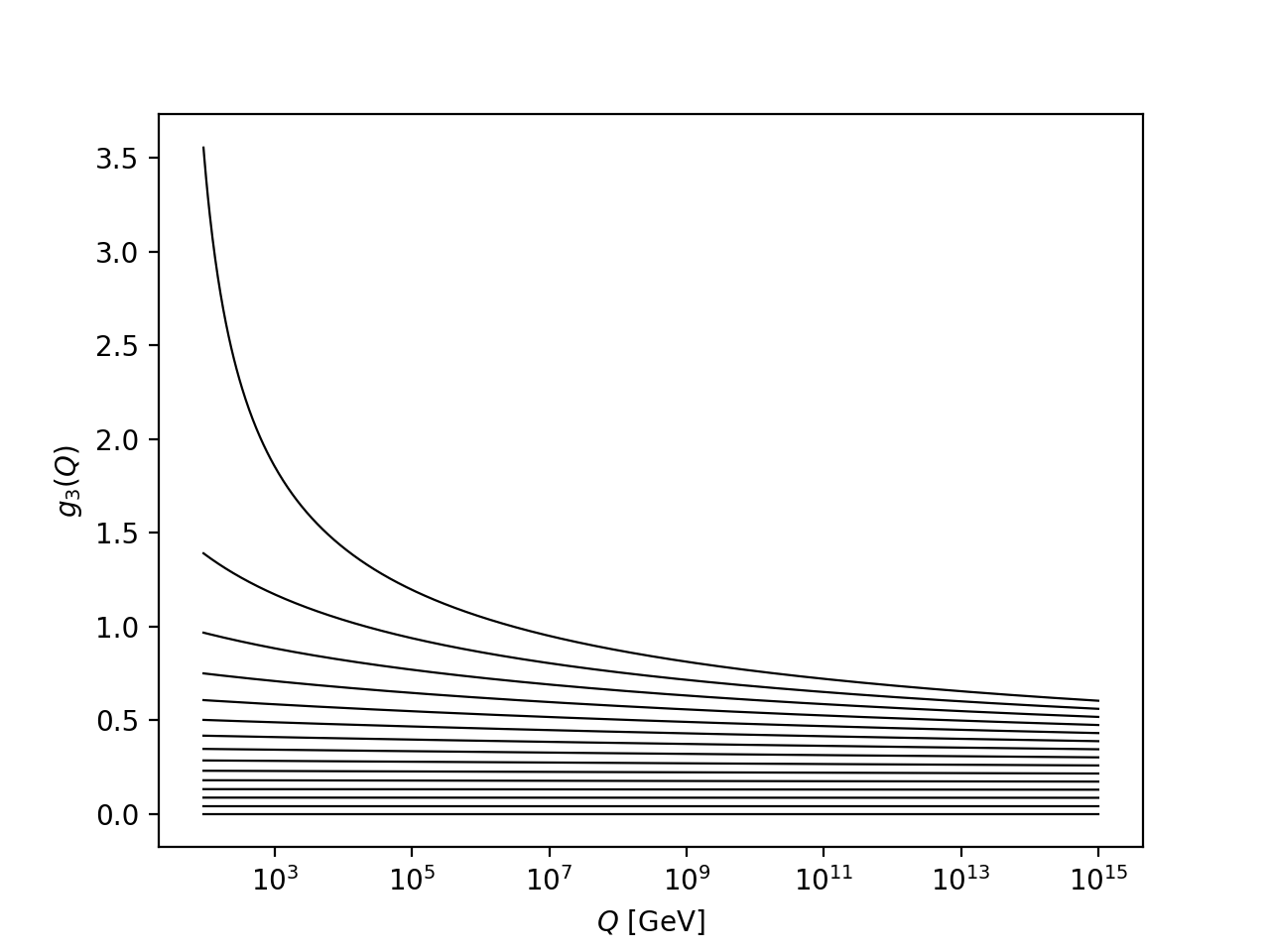} 
\caption{Probability flow lines for the $g_3(Q)$ gauge coupling evolved from $M_H=10^{15}\gev$ to $M_Z$. Equal spacing at $M_H$ indicates flat distribution (each value equally likely in the range), whereas the converging (diverging) of flow lines at $M_Z$ indicate increased (decreased) probability density of $g_3(M_Z)$ at low (high) values.}
\label{fig:g3rge}
\end{center}
\end{figure}

As we see from fig.~\ref{fig:fgmz} that $f(g_3(M_Z))$ peaks at low values of $g_3(M_Z)$. At first this may seem counter-intuitive, since the $g_3$ rises toward the infrared, and so should it not be more probable to have higher values? The answer is that $g_3=0$ is  a fixed point (albeit unstable) of the one-loop $\beta$ function and so low values of $g_3$ in the UV stay low in the IR whereas higher values of $g_3$ in the UV diverge rapidly in the IR. One can see this behavior by plotting probability flow lines for $g_3(Q)$, where evenly spaced values $g_3(M_H)$ are chosen  to reflect its flat distribution and then evolved down to low scales. See fig.~\ref{fig:g3rge}. The flow lines are denser for lower values of $g_3(Q)$ than at higher values of $g_3(M_Z)$. The density of flow lines at $M_Z$ is indicative of the probability distribution of $g_3(M_Z)$ at $M_Z$. Thus, there is higher probability for lower values.

How are we to interpret the flow of probability density, as defined above? It appears that nature's choice of $g_3$ appears to be more probable or less probable depending on what scale we evaluate it. A quantum field theorist might immediately recoil from this conclusion, since we are used to the maxim that observables (i.e., things that have meaning and a fixed value independent of how you might calculate them) cannot depend on what arbitrary scale you use to conduct perturbation theory. However, we are not computing observables, and so the maxim need not apply. Nevertheless, we are left asking what scale is most appropriate to ask about the local probability density of a coupling's value. As we discuss at the end of this section, the resolution to this question is that the scale does not matter if we specify a finite integration domain. In terms of values of the couplings, RG flow will expand and contract that finite domain  at different scales but the total probability within will remain fixed.

As with most probability discussions, it is fruitful to think of betting. If a distribution is given at the high scale for $g_3(M_H)$ and one is given a $\Delta g_3$ chip of some fixed finite range to place at the $M_Z$ scale, what choice of position $g_3(M_Z)$ would you put this chip? If one believes that that question makes sense and there is a computable answer, which appears to be so, then one might be convinced of probability flows of coupling distributions from RG evolution.

We note that in normal science, where $\delta$-function distributions are implicitly assumed for parameters, RG flow will retain $\delta$-function distribution centered on the coupling throughout its trajectory. Therefore, there is no conundrum to solve about what scale to evaluate a coupling's probability -- it is equally 100\% probable all throughout its RG flow.

Another implication of this discussion is that a flat probability distribution of a non-abelian gauge coupling from 0 to $\infty$ in the UV would push an infinite number of flow lines into a confining territory well before reaching our low value of $\Lambda_{\rm QCD}$ and thus our theory would be vanishingly improbable. Note, this conclusion could not be made by looking only at the high scale flat distribution, which says any value is equally likely. Only after RGE flow does one see that the non-infinite coupling density function (where $g_3(Q)<\sqrt{4\pi}$ by convention here) becomes infinitesimal after RG flow into the IR. The binary probability determination of what choices lead to too early confinement and what choices lead to QCD at $Q\leq \Lambda_{\rm QCD}$ is easy to make, and in that case the realization of QCD would have infinitesimal probability. For this reason it is somewhat safe to say that if there are probability distributions on QCD coupling they would not extend uniformly to very high values in the UV.

Repeating this exercise for a non-asymptotically free coupling, such as an abelian gauge coupling $e$, one finds that the probability distribution is more peaked at the higher values of the coupling in the IR given a flat distribution in the UV. As we flow deeper and deeper into the IR the probability density peaks further and further toward the maximum allowed value of the coupling as a function of scale.  Let us define $e_{\rm max}(Q)$ to be the maximum value of $e$ at the scale $Q$ given a maximum value of $e$ at $M_H$. If 
\beq
\label{eq:emaxQ}
\frac{de}{d\ln Q}=\frac{b}{2}e^3,~{\rm with}~b>0,~~{\rm then}~~e^2_{\rm max}(Q)=\frac{e^2_{\rm max}(M_H)}{1+e^2_{\rm max}(M_H)b\ln M_H/Q}.
\eeq
Also,
\beq
{\rm if}~~e_{\rm max}(M_H)=\infty~~~\Longrightarrow~~e^2_{\rm max}(Q)=\frac{1}{b\ln M_H/Q}.
\eeq
If $e(M_H)$ is flatly distributed from 0 to $e_{\rm max}(M_H)$ then the probability distribution for $e(Q)$ is
\beq
\label{eq:feQ}
f(e(Q))=\frac{1}{e_{\rm max}(Q)}\frac{(1-e^2_{\rm max}(Q)b\ln M_H/Q)^{1/2}}{(1-e^2(Q)b\ln M_H/Q)^{3/2}}.
\eeq
where $e_{\rm max}(Q)$ is given in eq.~\ref{eq:emaxQ}.

The first property to note of eq.~\ref{eq:feQ} is that if $e_{\rm max}(M_H)$ is indeed infinite the probability density function diverges at  $e(Q)=e_{\rm max}(Q)$. This is because all large RG flow lines, which were evenly spaced in $e(M_H)$ at $Q=M_H$, converge on $e_{\rm max}(Q)$. Only a relatively small number of lines (actually, infinitesimally small number of lines) converge to values discernably less than $e_{\rm max}(Q)$. This is because a small number of flow lines near $e(M_H)\sim 1$ are overwhelmed by the infinite number of flow lines that started from $e(M_H)\gg 1$. It is in this sense that we can call $e_{\rm max}(Q)$ a probability flow fixed point. An implication of this discussion is that we know that a distribution with infinitely many more flow lines for $e(M_H)\gg 1$ than for $e(M_H)\sim 1$, such as a flat distribution, could not have enabled $e<e_{\rm max}$ in the IR with any reasonable probability. Thus, there is likely a firm upper bound on the distribution of abelian couplings, which is the same conclusion we reached for non-abelian gauge couplings but for different reasons. Of course, the theory becomes non-perturbative as the couplings go higher, but the qualitative message is similar even if the value of the $e_{\rm max}(M_H)$ must remain below some agreed upon perturbative bound.

These same considerations lead one to conclude that $e_{\rm max}(Q)$ is a probabilistic fixed point deep in the IR independent of what value $e$ started with at $M_H$. To see this we note that the RG solution (one loop) is
\beq
e^2(Q)=\frac{e^2(M_H)}{1+e^2(M_H)b\ln M_H/Q}
\eeq
As $Q\to 0$ in the IR, the solution increasingly asymptotes to the $e(Q)\to 0$, but also asymptotes to $e_{\rm max}(Q)$. The proper way to analyze this is to ask for some initial choice of $e(M_H)$ below $e_{\rm max}(M_H)$ what value of does $e(Q)$ have with respect to $e_{\rm max}(Q)$. The answer is
\beq
\frac{e(Q)}{e_{\rm max}(Q)}=\frac{e(M_H)}{e_{\rm max}(M_H)}\sqrt{\frac{1+e_{\rm max}^2(M_H)b\ln M_H/Q}{1+e^2 (M_H)b\ln M_H/Q}}.
\eeq
At $Q=M_H$ we have $e(Q)/e_{\rm max}(Q)=e(M_H)/e_{\rm max}(M_H)$, but at $Q\to 0$ the expression asymptotes to $e(Q)/e_{\rm max}(Q)=1$. Thus, we can conclude that under high-scale flat distributions abelian theories asymptote in the IR to their maximum allowed values. This conclusion holds for many other distributions beyond flat.

To end this section, let us remark that if we have a density function $f(x)$ over a domain $a<x<b$, then we can always remap the random variable $x$ into $y$ by $y=\psi(x)$ such that the domain of $y$ is the minimum and maximum values of $\psi(x)$ over $x$'s domain $a<x<b$. By a suitable choice of $\psi(x)$ one can obtain a distribution function $f(y)$ that is arbitrarily large or small at $y_0=\psi(x_0)$. The implication for this is that knowing the  local value of a probability density function --- e.g., $f(x_0)$ ---  is not sufficient to  know how likely it is that the value of $x$ is $x_0$. We can only ask how likely it is to find $x$ in some finite domain $(x_1,x_2)$ of $x$-values enclosing $x_0$, and equivalently how likely it is to find $y$ in the corresponding mapped domain $(\psi(x_1),\psi(x_2))$. That is something one can place and win well-defined bets on, as discussed above. So, implicit to the above discussion is the assumption that we want to know how likely is it that the gauge couplings $g$ or $e$ falls within some small window, say $g=g_0\pm 0.1$, or what is the probability that the coupling is below some value $\hat g$. Those are meaningful questions that can be asked and answered in RG flows on parameter distributions at different any scale, where the function $\psi()$ is the analog to RG evolution.

\section{Fixed points and naturalness}

The investigation into probabilistic interpretations of theories has lead us toward fixed points. Indeed, the intuitions of researchers have always been that low-energy fixed points are the most probable values of those couplings. This is based on an implicit assumption of underlying flat distributions of parameters or distributions not too dissimilar from flat. Furthermore, IR fixed points have very low finetuning from standard finetuning functionals like those we discussed above. Very large changes in the UV (input parameters) yield  very small deviations in the IR (output parameters). However, one should caution that even in the presence of fixed points there are flow lines that deviate far from the fixed point values at some non-zero IR energy, and probability distributions can be made on the UV inputs that would favor those lines and disfavor the lesser finetuned values very close to the fixed point. Nevertheless, we can tentatively hold  that low finetuning may be indicative of higher probability for couplings, and therefore higher plausibility of a theory at that point in parameter space. Again, it must be emphasized that such a conclusion is based on an implicit assumption of underlying parameter distributions which are not too dissimilar to flat distributions, and therefore the hint of higher probability from lower finetuning is not guaranteed. 

\begin{figure}[t] 
\begin{center}
\includegraphics[width=0.6\textwidth]{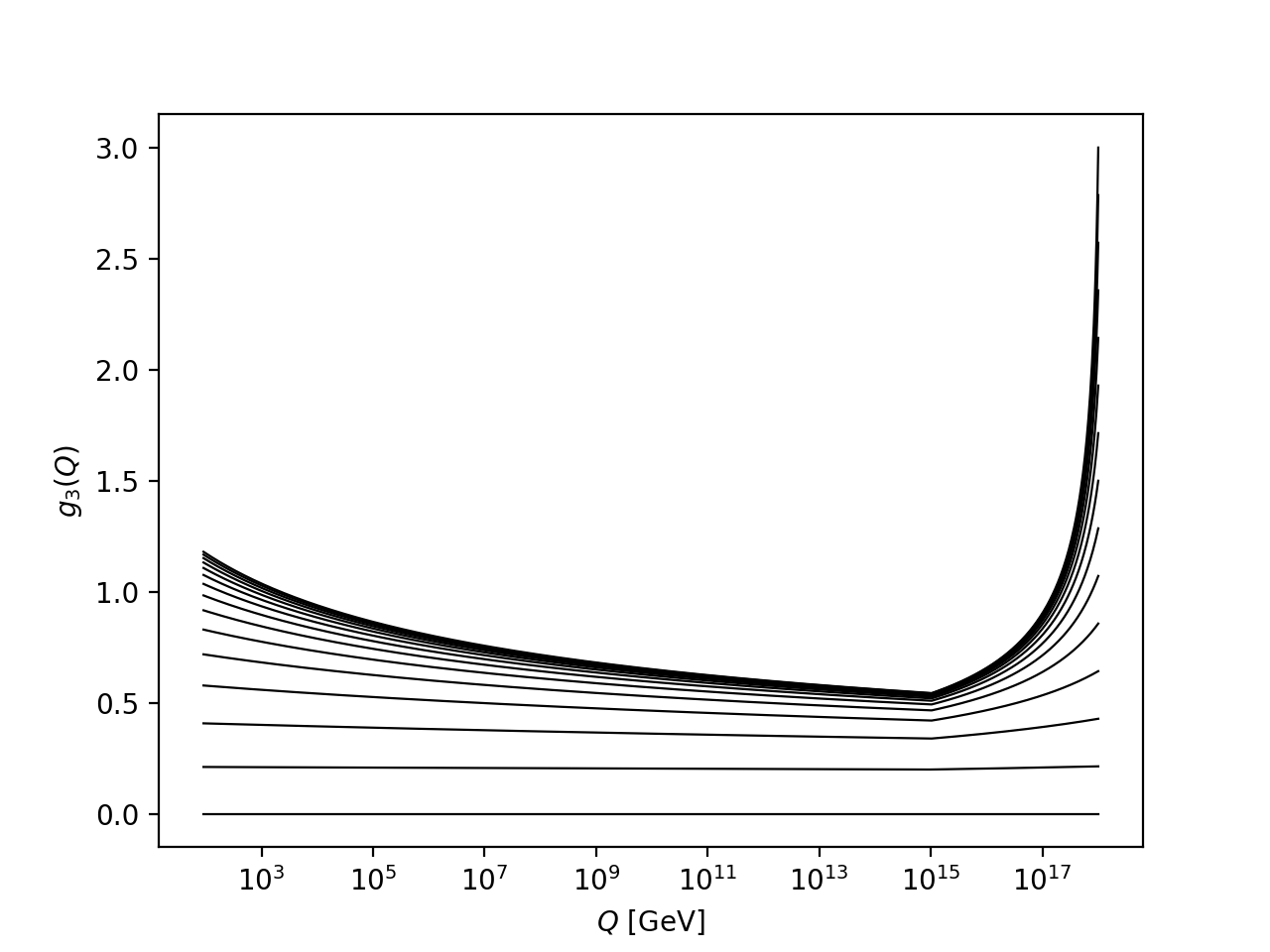} 
\caption{Probability flow lines for flatly distributed $g_3(Q)$ (unified coupling) at a scale of $2\times 10^{17}\gev$ evolved according to a strong non-asymptotically free GUT theory down to $10^{15}\gev$ and then evolved according to the Standard Model asymptotically free theory from $10^{15}\gev$ down to $M_Z$, yielding enhanced probability density at its maximum IR value. }
\label{fig:g3rgeFP}
\end{center}
\end{figure}

Probability fixed points of this nature may be used in an interesting way by nature. For example, the Standard Model may be unified into a grand unified theory (GUT) with large dimensional representations. In that case, the GUT would not be asymptotically free. If the GUT gauge coupling is flatly distributed at a very high scale it could yield a convergence of flow lines to an IR probability fixed point for the coupling. Here the IR refers to the GUT scale, where the GUT gauge group breaks to the SM model. The high density of flow lines reflects a large probability density. This large bunching of probability lines at the GUT scale then acts as inputs to the probability density for the SM gauge couplings.  The $SU(2)$ gauge couplings and especially the abelian $U(1)_Y$ gauge coupling squeeze these probability flow lines even closer together in the IR, peaking the distribution at a very narrow range.  The QCD gauge coupling, since it is asymptotically free, wants to fan the probability lines out and reduce the probability density along this $g_{\rm max}(Q)$ trajectory. However, if the original density function of $g_{\rm max}$ is sufficiently large at the GUT scale --- the flow lines being very packed there -- the fanning out process does not fully unravel the prediction and the low scale coupling is rather well peaked even for QCD. Fig.~\ref{fig:g3rgeFP} depicts the scenario discussed above, where flow lines are equally spaced at a scale well above the GUT scale, and then RG flow squeezes them closer together near the upper-limit quasi-fixed point. This could be an argument for GUT theories not being asymptotically free.

The above discussions on gauge coupling RG flows and probabilistic interpretations is simplistic from several points of view. First, it was all done in a one-loop analysis, which enabled us to see analytic formula for RG evolution and probability density functions. One should go to higher order in RG evolution, which in realistic theories of nature will involve additional couplings, such as the top Yukawa coupling and the Higgs coupling, albeit at suppressed order. Nevertheless, fixed point behaviors do change especially in strongly coupled regimes due to the existence of other couplings entering the flow. Once additional couplings are added the discussion of probability is greatly intensified. There are probability distributions that need to be assumed for all the parameters, and their correlations. A large joint probability function over all parameters is needed, in other words, to analyze further claims of likelihood of a theory point. A line of inquiry could be to ask what distributions (beyond the obvious $\delta$-function distributions) of high-scale parameters would yield RG flow lines that converge in the IR to the measured values. Such theories would then be natural by definition, and likely would involve little finetuning. However, since no theory has perfect flow to IR fixed point behavior, the preponderance of lines flowing to the IR fixed point neighborhood may still not be good enough for the strongest skeptic who would claim that an explicit probability distribution is needed to make any rigorous statement at all.

%%%%%%%%%%%%%%%%%%%%%%%%%%%%%%%%%%%
\section{Implications of skepticism}

The critical discussion above leads to a credible skepticism to extra-empirical theory assessments (SEETA). We do not have a meta-theory that provides statistical distributions to parameters of theories, which is a key problem precision naturalness arguments face.  It is therefore of some use to go through the exercise of promoting SEETA to a guiding principle, in apposition to naturalness and other extra-empirical assessments, to see what it might lead to. As we will see, the price of SEETA is rather steep compared to our ordinary intuitions about assessments of theories and their parameter spaces. Nevertheless, it is important to articulate these implications so as not to leave SEETA-like thinking nebulous, unexamined, and perhaps more attractive than it might otherwise be. 

 For us here, SEETA by definition is the disbelief that extra-empirical theory assessments, such as naturalness, enable one to find the ``{\it correct theory}," or the ``{\it more correct theory}" among competing theories. There are many resulting implications once SEETA is adopted, of which we highlight a few below.

First, since there is no meta-theory of probability distributions and any criteria to assess naturalness, such as finetuning measures, are unacceptable to SEETA, there can be no preferred regions in concordant theory space. Thus, any theory point is as good and {\it a priori} equally likely as another. For this reason, there can be allowed no disillusionment  of a theory even if experiment rules out a ``massive fraction" of allowed parameter space, since such declarations implicitly assume some knowledge of probability distributions of parameters, which, however, is barred from consideration by SEETA. Therefore, as long as there is at least one theory point that is surviving the theory is still as good as it ever was, and no judgments of reduced plausibility can be tolerated.

A second related implication of SEETA  is that only falsifiable theories allow their plausibility status to change, but only after discovery or after null experiments with total theory coverage. A falsifiable theory must enable the prospect of all its theory points to be ruled out by experiment. Here we include the condition that falsifiable theories make predictions across its entire parameter space that have not yet been confirmed by experiment. If the entire parameter space of the theory can be covered, then a falsifiable theory will either be ruled out because nothing new is found, or it is established beyond the standard theory since a non-trivial prediction was borne out by experiment. 

Now, the subtlety with this second implication is that one can extract out of any testable theory a falsifiable theory. For example, I may have a theory $T$ (e.g., supersymmetry), which has two regions of parameter space $T_{\rm F}$ and $T_{NF}$, where $T=T_F\oplus T_{\rm NF}$. One region, $T_F$, is the area that makes new predictions and can be non-trivially tested by experiment in the reasonable future (e.g., very low energy minimal supersymmetry). Another region, $T_{\rm NF}$, is the region of parameter space that is not $T_{\rm F}$.  If a theory $T$ has $T_{\rm F}\neq \emptyset$ then the theory is testable, and if $T_{\rm NF}=\emptyset$ it is falsifiable. If $T_{\rm NF}\neq \emptyset$ then one can declare a new theory $T'$ which is the projection $P_{\rm F}$ of the falsifiable region $T_{\rm F}$: 
\beq
T'=P_{\rm F}(T)={P}_{\rm F}(T_{\rm F}\oplus T_{\rm NF})=T_{\rm F}.
\eeq
$T'$ is a falsifiable theory, albeit artificially created, and its plausibility status is guaranteed to change after experimental inquiry. 
An example of this is projecting all of supersymmetry ($T$) down to a very minimal supersymmetric $SU(5)$ GUT with low-scale supersymmetry ($T_{\rm F}$), which is falsifiable and indeed was falsified~\cite{Murayama:2001ur}.

The difficulty with such falsifiable projections is the sometimes artificial nature of the division between $T_{\rm F}$ and $T_{\rm NF}$. The separation between the two is sometimes made not out of theory considerations, in contrast to the $SU(5)$ supersymmetric GUT example given above, but rather the perceived boundaries to what experiments are willing and able to achieve in the near term. For example, if $T$ is supersymmetry, and $T_{\rm F}$ is what the LHC can find, then there is a tendency to misname $T'=P_{\rm F}(T)=T_{\rm F}$ as ``supersymmetry", and when it is not found, it is said that ``supersymmetry" has been ruled out. In reality, $T'$ is better called ``LHC-projected supersymmetry", and therefore the LHC is capable of ruling out only ``LHC-projected supersymmetry", and not ``supersymmetry,"  if it is not found. Projecting $T$ onto $T_{\rm F}$ to form a falsifiable $T'$ based only on recognizing what experiment can and cannot do in the near term creates artificial theories whose falsification is not very meaningful.

Finally, a third implication of SEETA is that theory preference then becomes not about what theory is more likely to be correct but what theory is practically more advantageous or wanted for other reasons. Such reasons include fewer parameters, easier to calculate, has new experimental signatures to pursue, interesting connections to other subfields of physics, mathematical interest, etc. There are many reasons to prefer theories beyond one being more correct than another --- the only attitude unacceptable to SEETA is to say that one theory is more ``correct" or more likely to be correct than another theory that is equally empirically adequate. No theory of theory preference will be given here, except to say that ``diversity" has a strong claim to a quality for preference\footnote{Diversity is highlighted here, since its useful role has not been emphasized directly in the literature. Of course, many others extra-empirical preferences have strong claims too, such as a coherent and efficient explanation of causal mechanisms, etc.}.  If theorists only develop and analyze theories that give the same phenomena, at the expense of exploring other theories equally compatible with experiment, there becomes the practical problem of not arguing for or analyzing new signals requiring new experiments. A few examples out of many in the literature that have the quality of diversity at least going for it are clockwork theories~\cite{Giudice:2016yja,Giudice:2017fmj} and theories of superlight dark matter (see, e.g.,~\cite{Hochberg:2015pha,Hochberg:2015fth}). These theories lead to new experiments, or new experimental analyses, that may not have been performed otherwise. 

To conclude and summarize the skeptical ethos: theories must be compatible with experiment, and any that are should be viewed as just as likely as any others. There is no concept of a {\it speculative theory} being ``almost ruled out" when its parameter space shrinks due to null experimental results, since such descriptions imply knowledge of as-yet unknown probability distributions of parameters. Regarding naturalness, it becomes a theory quality upon a well-defined algorithm to quantify it. Without yet having a meta-theory of probability distributions over parameters, one must abandon standard naturalness as a quality that points to theories that are more likely to be correct, which may be fatal for naturalness since that is implicitly its main {\it raison d'\^etre}. It is important, nevertheless, to continue to assess theories beyond their empirical adequacy. Seeking diversity is one example of a potentially fruitful extra-empirical criterion. Other qualities such as simplicity, calculability, consilience, etc., may also be useful practical qualities to declare preferred theories. However, as argued, there is no known guaranteed justification to call these preferred theories more correct.

%%%%%%%%%%%%%%%%%%%%%%%%%%%%%%%%%%%%%%%%
\section{Summary: three positions on naturalness and finetuning}

In the way of summary, three viewpoints that have been coursing through the discussion will now be directly expressed, and a brief  appraisal of their viabilities will be given in light of the above discussion.

First, there is the ``extreme pro-naturalness position," which holds that all correct theories, at all times and from all levels of understanding, and from all perspectives no matter how limited, must be natural and non-finetuned. One can at times even make precision statements about what numerical values parameters and observables must be bounded by in order to satisfy the rigid demands of naturalness, such as was even done for superpartner masses by the original precision finetuning papers of Anderson \& Castano~\cite{Anderson:1994dz,Anderson:1994tr}.  The discussion above has given plenty of evidence against the surety of such an extreme position.

Second, there is the ``extreme anti-naturalness position," which holds that there is no role whatsoever in favoring or disfavoring a theory based on finetuning or naturalness assessments. Finetuned theories at any level of finetuning are just as likely as any other theory. The discussion above does not prove or even advocate this viewpoint; however, it has demonstrated that the position is not wholly unreasonable given our current understanding. The literature has not sufficiently articulated an incontrovertible argument against it, and it remains an interesting philosophical viewpoint that needs further careful and devoted discussion. 

The implications of the ``extreme anti-naturalness position" are severe, and are useful to identify. They are summarized by the SEETA implications discussed in the previous section. For example, as a very specific illustration, one cannot hold an extreme anti-naturalness position and simultaneously declare that low-energy supersymmetry (e.g., all superpartners below 1 TeV) is ruled out by the LHC when there exists even one point in low-energy supersymmetric parameter space that survives LHC experimental probes, which is currently the case\footnote{For example, all superpartners are degenerate in mass at 1 TeV.}. Other implications of SEETA violate first-blush sensibilities of demarcating good vs.\ bad theories, but such sensibilities relevant to naturalness need to be articulated much more precisely in the literature to eliminate the SEETA viewpoint, if it is possible to do so.

Finally, there is the ``moderate naturalness position", which holds that,
generally speaking, theories are not finetuned and they are natural, even though perhaps there is a small fraction of cases where there is some very high tuning. Such rare cases, however, do not vitiate the utility of naturalness, just as the rare case of a $B$ meson decaying to a strange meson does not eliminate the understanding that a $B$ meson decays much more often to a charm meson.  Thus, theories and theory points that appear unnatural and finetuned are generally not really unnatural and finetuned once we have understanding of a deeper theory from whence the current theory derives. The discussion above leaves this viewpoint credible also. 

An important implication of ``moderate naturalness position" is that the search for natural theories is a valid enterprise. It also implies, on the other hand, that finetuned theories from our current perspective are not necessarily wrong theories. They could be that way by rare accident of nature or are seen to not be finetuned from a deeper/different perspective. The moderate position has a difficult time with questions of ``extreme finetuning," since it implies that it can never be accidental and new principle(s) must be found to explain it. ``Finetuning that cannot be accidental"  is perhaps the best definition of ``extreme finetuning", and it becomes a research question, not resolved here, how to demarcate between extreme and merely large finetunings, if a boundary does indeed exist. 

Not every view of naturalness and finetuning is fully captured by one of the three primary positions described above. Nevertheless, they do reflect qualitative attitudes that have existed in the literature reasonably accurately. Perhaps the single most important qualitative viewpoint emerging from this work is that fervid positions on the utility or the non-utility of naturalness and finetuning arguments have much more work to do to prove their indisputable merit. There are plenty of disquietudes presented against both extremes, which warrants careful consideration and tentative adoption of the ``moderate naturalness position."

\medskip\noindent
{\it Acknowledgments:} I am grateful for discussions with A.~Hebecker, S.~Martin, A.~Pierce, and Y.~Zhao. This work was supported in part by the DOE under grant DE-SC0007859. 

%\vfill\eject

%%%%%%%%%%%%%%%%%%%%%%%%%%%%%%%%%%%

\end{document}